# Monitoring of tritium purity during long-term circulation in the KATRIN test experiment LOOPINO using laser Raman spectroscopy

S. Fischer[a], M. Sturm[a,*], M. Schlösser[a], B. Bornschein[a], G. Drexlin[b], F. Priester[b], R.J. Lewis[c], H.H. Telle[c]

[a] *Institute for Technical Physics, Tritium Laboratory Karlsruhe,*
*Karlsruhe Institute of Technology, 76021 Karlsruhe, Germany*

[b] *Institute for Experimental Nuclear Physics,*
*Karlsruhe Institute of Technology, 76021 Karlsruhe, Germany*

[c] *Department of Physics, Swansea University, Singleton Park, Swansea, SA2 8PP, United Kingdom*

*The gas circulation loop LOOPINO has been set up and commissioned at Tritium Laboratory Karlsruhe (TLK) to perform Raman measurements of circulating tritium mixtures under conditions similar to the inner loop system of the neutrino-mass experiment KATRIN, which is currently under construction. A custom-made interface is used to connect the tritium containing measurement cell, located inside a glove box, with the Raman setup standing on the outside. A tritium sample (purity > 95 %, 20 kPa total pressure) was circulated in LOOPINO for more than three weeks with a total throughput of 770 g of tritium. Compositional changes in the sample and the formation of tritiated and deuterated methanes $CT_{4-n}X_n$ (X=H,D; n=0,1) were observed. Both effects are caused by hydrogen isotope exchange reactions and gas–wall interactions, due to tritium β decay. A precision of 0.1 % was achieved for the monitoring of the $T_2$ $Q_1$-branch, which fulfils the requirements for the KATRIN experiment and demonstrates the feasibility of high-precision Raman measurements with tritium inside a glove box.*

## I. INTRODUCTION

Possible solutions for precise compositional analysis, in-line and in (near) real-time, of flowing gases containing tritium have now been discussed for more than a decade. Key applications are at present encountered in the fundamental research KATRIN (the Karlsruhe Tritium Neutrino mass experiment) collaboration, and in the various analytical systems for the Tritium Plant of the ITER (International Thermonuclear Experimental Reactor) project. In-line analysis of flowing gases is problematic at the best of times but it constitutes a major challenge if radioactive (or toxic) compounds are involved, for which the integrity of a circulation loop cannot be compromised.

Of the laser spectroscopic techniques now in routine use for compositional analysis of gases Laser Raman spectroscopy (LARA) has been identified as the method of choice for this task; it is the most suitable to cover a wide variety of expected molecular compounds and requiring only a single, fixed wavelength laser excitations source. A number of implementations were proposed, and promising results have been reported, albeit mostly for non-tritiated mixtures. [1-4]

The requirements for the KATRIN experiment are that the composition of the tritium gas (of purity >95%), which will be injected into its windowless gaseous tritium source (WGTS), needs to be known to a precision of 0.1% (1σ). Monitoring data have to be provided at least every 250 seconds. [5] In an experiment in which the conditions of the full KATRIN inner loop were mimicked (Test of Inner-loop setup, TILO), [6] we showed that we could meet these requirements for mixtures of hydrogen and deuterium, measured at the position of the buffer vessel in the KATRIN loop (for gas mixtures of total pressure 10-20 kPa). [4] More recently, we demonstrated that with our LARA system we can indeed also achieve this for gas mixtures containing all six hydrogen isotopologues, [7] although this proof-of-principle was done in a static-gas cell.

We are now at the stage of carrying out tests to determine, whether our Raman measurement system is suitable for the KATRIN inner loop with circulating tritium gas. Since not all components of the full KATRIN tritium loop are yet in place (the WGTS will become available only toward the end of 2011), this was done in a simpler loop, LOOPINO, which runs under KATRIN-conditions regarding throughput, pressure and composition. The purpose of these measurements was to ascertain the stability and reliability of the setup, both of the loop itself and of the Laser Raman system during runs over extended periods of time. Such measurements have proven to be crucial during the TILO test phase, [6] affecting some of the design issues of the KATRIN inner loop design. Also, LOOPINO will allow us to verify that reliable LARA measurements are possible inside the secondary containment of the glove box.

---

\* Corresponding author.
E-Mail address: michael.sturm @ kit.edu

In this paper, we describe and discuss results from our first long-time LARA measurements of tritium flows in LOOPINO, under KATRIN-like conditions.

## II. EXPERIMENT DESCRIPTION

### II.A. Technical requirements

Laser Raman measurements on tritium impose stringent technical requirements on the experimental setup concerning leak tightness of the setup, and radiation safety. All components of the primary system, i.e. those which are in direct contact to tritium, have to be made of certified material, metal sealed (typically stainless steel tubes, VCR- or CF-connections) and leak-tested to $< 10^{-9}$ mbar l/s for each individual connection and $< 10^{-8}$ mbar l/s integral leak rate.[8] When working with activities $> 10^{10}$ Bq, the primary system has to be enclosed by a secondary containment with tested integral leak rate of $< 0.1$ Vol.% h$^{-1}$. As much of the optical equipment as possible should be located outside the secondary containment to minimize the risk of contamination, and to allow for easy handling. Therefore, the connection between the glove box and the optical setup plays a major role in the design of the experimental setup.

### II.B. Experimental setup

Figure 1 shows the conceptual flow-diagram of the Inner-Loop system of KATRIN, which will be used to continuously feed the KATRIN WGTS with a constant rate of high-purity tritium gas (tritium purity $> 95$ %). A palladium membrane filter (permeator) is used to separate non-hydrogen compounds from the gas stream, e.g. the decay product $^3$He and $CT_{4-n}X_n$ (X=H,D) methane isotopologues.[5] The pressure within the pressure-controlled buffer vessel is stabilised via a PID-controller and a regulating valve. The Laser Raman sample cell (LARA cell) will be located between the two buffer vessels and used to monitor the gas composition in-line.

Since the Inner Loop system will not be operated with tritium until the commissioning of the WGTS a simplified circulation test-loop, LOOPINO, has been set up and commissioned. It consists of a buffer vessel (volume 1000 cm$^3$), the LARA cell and a pumping unit (Metal Bellows 601DC). The pressure in the buffer vessel, and hence also the pressure in the LARA cell, is controlled by a regulating valve and a PID-controller, while a coarse setting of the flow rate can be done by a needle valve. Since a permeator is not installed in this minimised loop, an enrichment of impurities during long term circulation is expected.

Figure 2 shows the glove box in which parts of the Inner Loop system as well as the Isotope Separation System (ISS)[9] of TLK and LOOPINO are located. The complete optical setup is located outside the glove box on a mobile optical table in a light-tight housing, called LARA setup. An appendix has been attached to the glove box at the interface between loop system and the LARA setup (see Fig. 3). Laser windows in the walls of the appendix allow the laser beam to pass through the LARA cell, which is connected to the circulation loop via bellows. The scattered light leaves the cell at 90° to the direction of excitation, through a third window in the appendix wall. The LARA cell is mounted on a suspended base plate for the decoupling of vibrations produced by pumps in the glove box. The optical path inside the appendix is enclosed by light-tight tubes to prevent ambient light to enter along with the Raman scattered light, and to ensure laser safety. Once moved to the right place, the LARA setup is screwed down, ensuring a fixed position of the LARA cell relative to the optical setup.

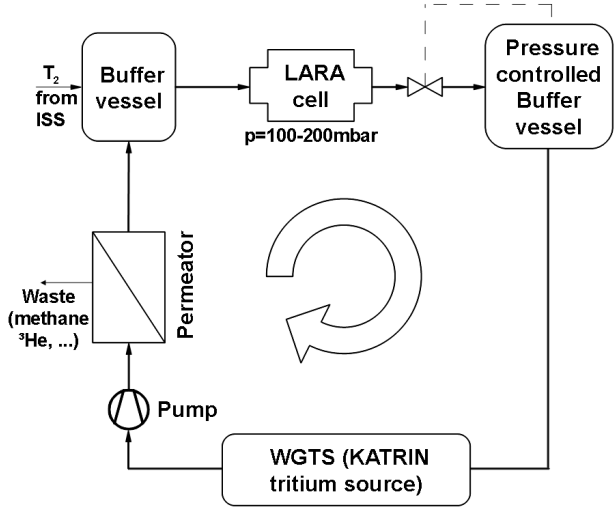

Fig. 1: Conceptual flow-diagram of the Inner-Loop of KATRIN. Note that LOOPINO is a simplified setup test version, excluding the buffer vessel, the permeator and the KATRIN tritium source.

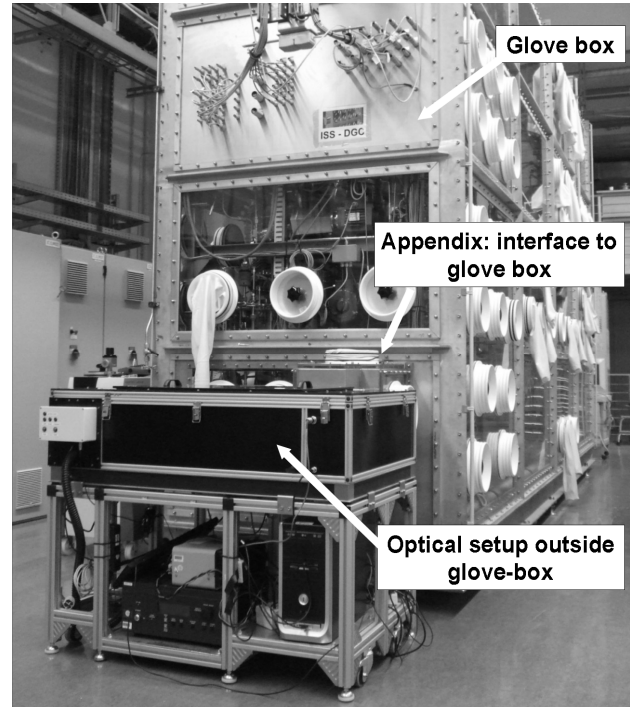

Fig. 2: LARA setup attached to the glove box. The appendix connects the optical setup outside the glove box with the LARA cell located inside.

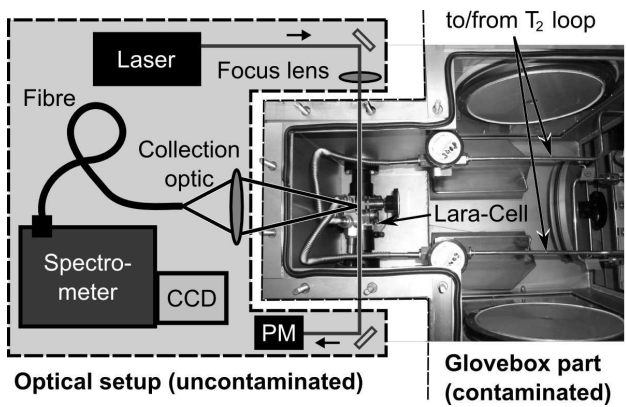

Fig. 3: Sketch of the connection of the LARA system (left) and the LARA cell inside the glove box, using the appendix (right).

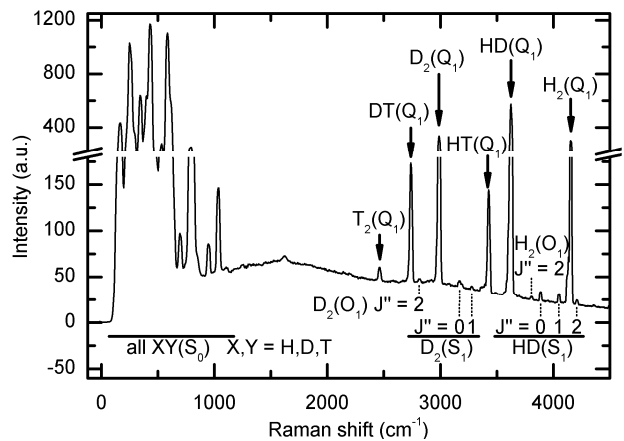

Fig. 4: Spectrum of lowly-tritiated gas sample with all six hydrogen isotopologues measured inside the appendix.

The LARA setup uses a 532 nm laser (Coherent *Verdi V5*) as the excitation source. After leaving the appendix, the scattered light is focused by a set of lenses onto an optical fibre bundle that transport the light to the spectrograph. A low-noise CCD (PI *PIXIS:2K*) is used to acquire the spectra. The LARA setup and the design of the LARA cell have been discussed in detail in previous publications [2,7], while a general discussion of the design of LARA systems for tritium monitoring can be found in Schlösser *et al*. [11] Automated software post-processing routines transform the raw 2D-image of the CCD into a spectrum. These routines include an astigmatism correction as well as a cosmic ray removal, which is necessary due to the long exposure times. [4]

Measurements at LOOPINO are typically performed in three steps:
(1) LOOPINO is evacuated to less than 10 Pa and afterwards filled with gas from the ISS and/or the CAPER facility [10] of the Tritium Laboratory Karlsruhe (TLK).
(2) The pumping unit of LOOPINO circulates the gas for several weeks through the LARA-cell and the buffer vessel, while Raman measurements (250 s acquisition time, 5 W laser power) are continuously performed.
(3) After a measurement run the gas is being transferred back to the CAPER facility for tritium-recovery.

The CAPER facility can be also used to perform gas chromatographic (GC) analyses of tritiated gas mixtures with about 5 % relative uncertainty.

### III. RESULTS AND DISCUSSION

### III.A. Gas mixture with low levels of tritium

The first test-measurements at LOOPINO were performed with non-tritiated gas mixtures, e.g. with deuterium (not discussed here), and a gas mixtures with a low concentration of tritium that contained all six hydrogen isotopologues $T_2$ : DT : $D_2$ : HT : HD : $H_2$ (approximate concentration 0.9 : 8.3 : 16.3 : 9.0 : 4.6 : 31.8, in %, according to GC analysis). Figure 4 shows an example spectrum of this mixture, which was circulated in LOOPINO for about 160 hours. The $Q_1$-branches of the hydrogen isotopologues, which correspond to purely vibrational excitations ($\Delta v = 1$, $\Delta J = 0$), are in the spectral range 2450–4250 cm$^{-1}$ and are well separated. The $S_0$-branches, related to pure rotational excitations ($\Delta v = 0$, $\Delta J = 2$), are found in the first 1500 cm$^{-1}$ of the spectrum and cannot be fully resolved with our combination of spectrograph and CCD.

The underlying background spectrum consists of the Lorentz wing of the laser line (strongly suppressed by an edge filter) as well of Raman scattering and fluorescence from the SiO$_2$ cell windows (peaking at ~450 cm$^{-1}$). [7] Due to these two reasons, all quantitative analysis is based on the $Q_1$-branches of the hydrogen isotopologues. Besides the $Q_1$-branches, several weaker spectral lines of the main gas constituents $D_2$, HD and $H_2$ can be observed, namely $S_1$ ($\Delta v = 1$, $\Delta J = 2$) and $O_1$ ($\Delta v = 1$, $\Delta J = -2$) lines.

Besides the new gas composition no significant differences between the spectra of the circulated mixture and of a static sample [7] were found, i.e. the additional windows in the appendix and the vibrations from the pumping units inside the glove box did not influence the measurements.

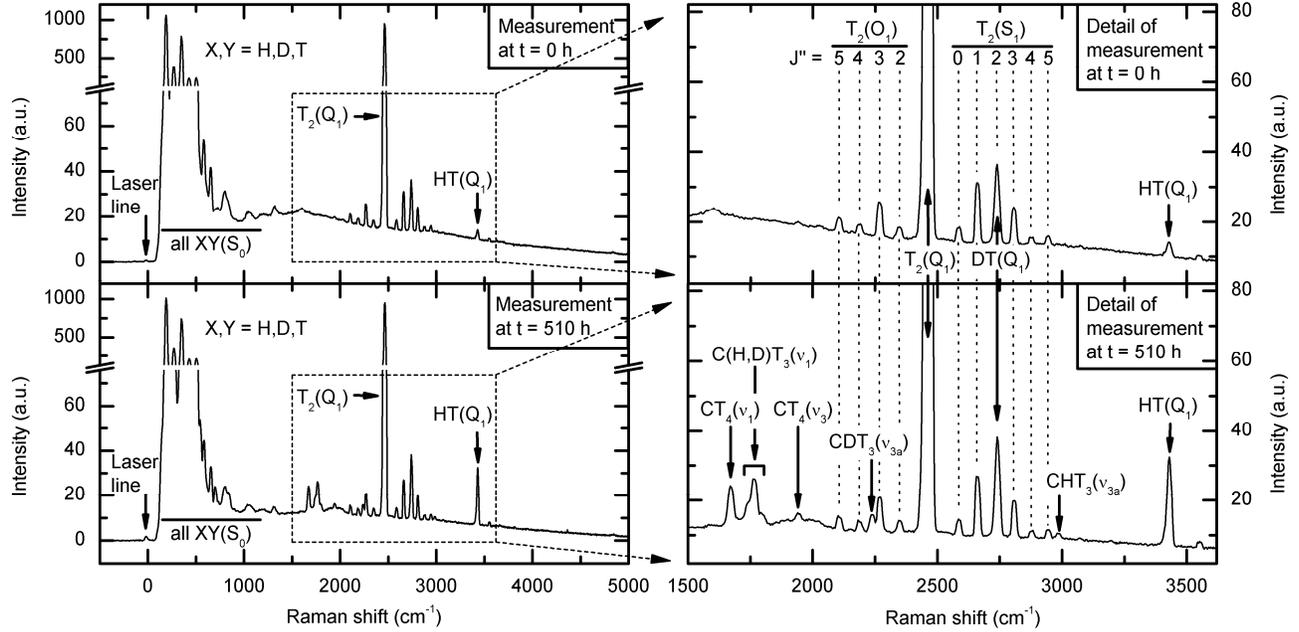

Fig. 5: Spectra at the start and at end of the measurement run of high purity tritium mixture. Note the overlap of the DT ($Q_1$) and $T_2$ ($S_1$, J''=2) peaks indicated in the enlarged sections on the right. Both spectra are normalised to the $T_2$ ($Q_1$) peak.

The absolute peak intensities of the $Q_1$-branches of the hydrogen isotopologues varied during the monitoring by up to 5 %, due to laser power fluctuations. In order to be insensitive to such laser power fluctuations, relative peak intensities were used for the analysis, i.e. the intensity of each $Q_1$-branch in the spectrum is normalized on the sum of all $Q_1$-branch intensities. The relative peak intensities varied only within 0.5 % since all absolute peak intensities are equally affected by laser power variations. Therefore, relative peak intensities seem to be suitable when monitoring gas mixtures with known constituents.[7] In a first approximation, the relative intensities are interpreted as concentration ratios. Earlier measurements showed that this approximation is valid within the stated tolerances of commercial non-active gas mixtures, and within the GC measurement uncertainties for tritiated mixtures. This approximation is used in the following.

A more detailed analysis is possible when one incorporates the individual transition probabilities of spectral lines, the fourth-power wave number dependence of the scattering intensity [12] and the quantum efficiency of the collection optics. All these corrections will be implemented in the analyses in the near future. Calculated transition probabilities are available for all hydrogen isotopologues.[13,14] Experimental data, in contrast, are only found for the non-tritiated species.[15,16] Hence, measurements for the tritiated isotopologues are under way at TLK.

### III.B. Gas mixture with high levels of tritium

After these first measurements, LOOPINO was evacuated and filled with 18.3 kPa tritium gas (purity > 98 %) which corresponds to an inventory of about 95 mg tritium and $3.4 \cdot 10^{13}$ Bq respectively. A gas chromatographic measurement, performed before the filling process, detected as main constituent about 99 % $T_2$, traces of DT and HT with concentrations below 1 %, while $D_2$, HD and $H_2$ were not detected. After starting the gas circulation the pressure stabilised to $20 \pm 0.03$ kPa within about 5 minutes. The circulation was interrupted after 217 hours of continuous operation for about 23 hours, due to a power failure in the glove box. The LARA measurements were not affected, and hence were continued without interruption. No change in the gas composition and no impact on the LARA setup were observed. After power restoration, the stabilised circulation and LARA monitoring was continued for a further 310 hours. In total, a gas throughput of $99.6 \pm 1.1$ sccm was established over more than 500 hours, which corresponds to $(1.57 \pm 0.02) \times 10^{11}$ Bq/s $(4.25 \pm 0.05$ Ci/s) within this measurement run. In total about $2.8 \times 10^{17}$ Bq (7.7 MCi), i.e. more than 770 g tritium were circulated through the LARA cell.

Figure 5 shows two Raman spectra taken at the start and the end of the circulation period, respectively. The result of the gas chromatographic measurement is confirmed by the first Raman spectrum: The $T_2$ $Q_1$-branch dominates the spectrum, with about 97 % relative intensity; the relative intensities of the $Q_1$-branches of DT and HT are about 2.25 % and 0.5 %, while all other $Q_1$-branches have relative intensities of 0.15 %, or less.

The apparent discrepancy of the DT content between the GC analysis and the relative intensities in the Raman spectrum is due to the known overlap of the DT $Q_1$- branch and the $T_2$ ($S_1$, J''=2) peak.[7] The overlap was not considered in the analysis of the spectra, which thus results in an overestimate of the peak intensity for the DT $Q_1$-branch.

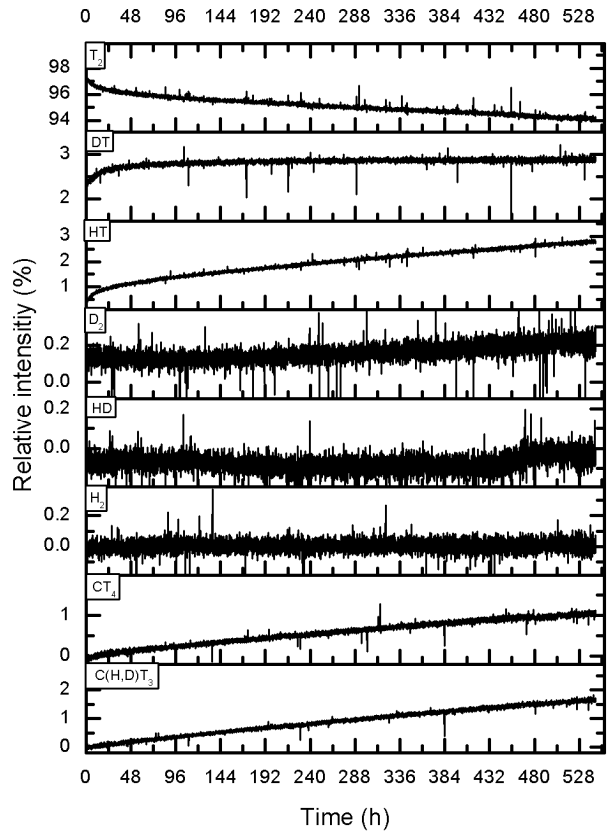

Fig. 6: Time evolution of relative intensities of hydrogen isotopologues and tritiated methane species in the sample.

During the circulation period the laser power decreased by about 20 %, and in unison so did all absolute peak intensities. The variations can mainly be attributed to thermal influences on the laser. For clarity, and to allow for easier comparison, all spectra in Fig. 5 were normalized to the peak intensity of the $T_2$ $Q_1$-branch, in order to eliminate the effect of decreasing laser power.

As in the case of the mixtures with low levels of tritium only minor changes of the relative concentration of the non-tritiated isotopologues could be observed (see Fig. 6). In contrast, the concentrations of the tritiated species varied strongly. The relative intensity of the $T_2$ $Q_1$-branch decreased to about 94 %, while the relative intensities of the DT and HT peaks continuously increased to about 3 %. The expected reduction of $T_2$ concentration in the sample, due to β-decay, is below 0.5 % and thus not sufficient to explain the observed change in composition. In addition, further features in the spectral range 1600–1800 cm$^{-1}$, and around 2250 cm$^{-1}$ and 3000 cm$^{-1}$ slowly appeared during the long-term monitoring. These features can be correlated to vibrational excita-tions of tritiated methane species, such as $CT_4$, $CDT_3$ and $CHT_3$.[17] The $\nu_1$-mode peaks associated with $CHT_3$ and $CDT_3$ at around 1750 cm$^{-1}$ could not be resolved individually with our spectrograph and CCD combination, and hence are jointly treated in the following. The methane $\nu_1$-mode peaks were included in the calculation of relative intensities, since not merely hydrogen isotopologues were present in the gas mixture.

The observed compositional changes are caused by hydrogen isotope exchange reactions and gas–wall interactions with the stainless steel tube walls, which both are triggered by the released energy from tritium β-decay.[18-21] Both effects will be discussed in more detail in the following paragraphs.

The time evolution of the $T_2$, DT and HT concentrations is described reasonably well by a double-exponential $f(t) = y_0 + A \cdot e^{-t/\tau_1} + B \cdot e^{-t/\tau_2}$ trend, with a common short time constant $\tau_1 \approx 11$-22 h and different, longer time constants $\tau_2 \approx 100$ h (for DT), $\tau_2 \approx 1000$ h (for HT) and $\tau_2 \approx 2000$ (for $T_2$). The short time constant $\tau_1$ is compatible with measurements of Uda et al,[18] who studied hydrogen isotope exchange reactions in glass cells. The longer time constants $\tau_2$ are more likely related to gas – wall interactions than to gas – gas interactions. Since $H_2$ had been the main constituent of most gas samples in LOOPINO before this measurement, a non-negligible amount of hydrogen atoms was dissolved in the tube walls. After filling LOOOPINO with the high purity tritium mixture the hydrogen atoms started to diffuse back into the gas volume, while tritium atoms were dissolved in the walls. This process did not stop within the measurement period due to the evidently rather large amount of hydrogen atoms in the tube walls and the high tritium concentration in the gas. Since LOOPINO was also operated with deuterium in prior test measurements, the formation of DT is very likely based on the same mechanism, but with a smaller initial deuterium concentration in the tube walls. The formation of DT therefore slowed down after 2-3 days due to the decreasing concentration of deuterium atoms in the tube walls, which yields to a significant smaller time constant $\tau_2$ for DT than for HT.

The concentrations of the tritiated methane species increase nearly linearly within the measurement period. The formation of tritiated methane species in presence of tritium gas and stainless steel was demonstrated by Morris,[21] who tested the influence of surface cleanliness of test vessels (volume 293 cm$^3$, 353 cm$^2$ inner surface) on the methane formation rate. For about 70 % of all investigated samples Morris found that, saturation of the methane concentrations occurred after 20-80 hours. The absence of saturation effects in our measurement is not surprising when taking into account the ×50 factor larger surface-to-volume ratio of LOOPINO (volume about 2000 cm$^3$, 1.2×10$^5$ cm$^2$ inner surface).

The achieved measurement precision $\sigma_I / I$, where $\sigma_I$ is the standard deviation of the residuals of the double-exponential fit and $I$ is the intensity of the peak of interest, can be calculated for each branch. A precision of 0.1 % is achieved over the whole measurement period for the $T_2$ $Q_1$-branch. This fulfils the requirement for LARA for the KATRIN experiment.[5] The precision of the second-most intense DT $Q_1$-branch (relative intensity < 3 %) is in the range 1.47–1.87 %; here the reduction in precision due to the decrease of laser power is quite visible.

The gas mixture was kept in LOOPINO after the circulation period for additional measurements. After three months of exposure to the highly-concentrated tritium mixture damage of the anti-reflection coatings on the inner side of all windows of the LARA cell was observed. Further investigations of this tritium-induced coating damage are ongoing.

## IV. SUMMARY & CONCLUSIONS

LOOPINO has been successfully commissioned and was operated the first time with a high purity tritium sample (95 mg tritium, $3.4 \cdot 10^{13}$ Bq) over a period of about three weeks. The feasibility of monitoring the gas composition based on Raman spectroscopy, using the appendix as connection between the optical setup and the tritium containing LARA cell, has been proven. In comparison to the measurements performed with static gas samples outside the glove box [7] no further serious disturbances have been observed. The experimentally achieved measurement precision of 0.1 % for the $T_2$ $Q_1$-branch fulfils the requirements for KATRIN experiment.

Hydrogen isotope exchange effects and the formation of tritiated methane species, due to gas – wall interactions, have been monitored in the highly-concentrated tritium mixture; tritium β-decay has been identified as trigger mechanism. In the lowly-tritiated sample the effect of exchange reactions is nearly unnoticeable. These findings are broadly in agreement with the study by Morris.[21] Although the permeator in the inner loop system of KATRIN will separate impurities from the gas flow towards the WGTS, further investigations on methane formation with LOOPINO are planned.

While having reached the KATRIN requirements, further improvements of the LARA setup towards improved sensitivity and precision are scheduled. In particular, the installation of a new temperature-stabilised 5 W, 532 nm laser will further improve the signal-to-noise ratio in the spectra, and accordingly the measurement precision. Also, the incorporation of a double-pass geometry has been successfully demonstrated in first test experiments at TLK. This nearly doubled the laser irradiance in the LARA cell by back-reflecting the laser beam for a second pass through the LARA cell.

## ACKNOWLEDGMENTS

This work has been partially supported by funds of the DFG (SFB/Transregio 27 "Neutrinos and Beyond"). The authors also would like to thank the TLK infrastructure team for their invaluable help.